\documentclass[a4paper,twocolumn, nofootinbib,superscriptaddress,aps,prl, 14pt,notitlepage,showkeys]{revtex4-1}
\usepackage{multirow}
\usepackage{graphicx}
\usepackage{bm}
\usepackage{dcolumn}
\usepackage{hyperref}
\usepackage{gensymb}
\usepackage{amsmath}
\usepackage{dcolumn}    
\usepackage{ifpdf}
\usepackage{amssymb,lineno,amsfonts}
\usepackage{graphicx}   
\usepackage{bm}         
\usepackage{bbm}
\usepackage{mathrsfs}
\usepackage{upgreek}
\usepackage{mathtools}
\usepackage{epstopdf}
\usepackage{setspace}
\usepackage{hyperref}
\usepackage{natbib}
\usepackage{esvect}
\usepackage{amsmath}
\usepackage{braket}
\usepackage[usenames,dvipsnames]{xcolor}
\definecolor{med-blue}{RGB}{25,25,112}
\hypersetup{colorlinks, linkcolor={blue},citecolor={blue}, urlcolor={blue}}
\usepackage{times }
\usepackage [english]{babel}
\usepackage [autostyle, english = american]{csquotes}
\MakeOuterQuote{"}
\begin{document}
	\title{Nonlocal drag by topological surface magnons in a pyrochlore ferromagnet}
	\author{Avirup De}
	\affiliation{Department of Physics, Indian Institute of Science Education and Research, Pune, India}
	\author{Dharmalingam Prabhakaran}
	\affiliation{Clarendon Laboratory, Department of Physics, University of Oxford, UK}
	\author{Sunil Nair}
	\affiliation{Department of Physics, Indian Institute of Science Education and Research, Pune, India}
	\date{\today}
	\begin{abstract}
The nontrivial topology of quasiparticle wavefunctions can manifest themselves in the form of observable surface states. This is now well established in electronic systems, with Dirac and Weyl semimetals bringing to fore the exotic nature of these topologically protected entities. Magnons - which refer to collective excitations of localized spins - offer another sector where many of these concepts could be realized. Here, we report magneto-thermal measurements on a pyrochlore ferromagnet which is theoretically predicted to host such topological magnons. It is demonstrated that the thermoelectric potential across a metal layer deposited on single crystalline specimens of Y$_2$V$_2$O$_7$ can be used to measure the magnon Hall effect. Moreover, a direct manifestation of topologically protected magnon surface states is observed - via the interfacial drag which these surface spin currents impose on the conduction electrons of the adjacent metallic layer.
	\end{abstract}        
	\pacs{Pacs}
	\maketitle

The electronic and thermodynamic properties of crystalline solids are often driven by wave-like collective excitations of quasi-particles associated with the lattice, spin and charge degrees of freedom. A spin wave is one such entity within a magnet, and corresponds to a wave-like excitation of precessing spins within an ordered magnetic background. The quanta of these spin waves - or magnons- are characterized by a wave vector ($\vec{k}$) and frequency ($\nu$), and the non-trivial topology of the magnon band structure could provide a bosonic parallel of the oft-encountered electronic one \cite{RevModPhys.90.015001}. However, topologically protected magnon surface states remain to be experimentally verified, owing to the limited sensitivity of most measurement techniques to these states \cite{mcclarty2021topological}. The topological attributes of these magnon bands is expected to manifest in heat and spin transport, in analogy to the spectroscopy and transport signatures observed in electronic systems \cite{RevModPhys.88.035005}. The most significant among them is the magnon Hall effect (MHE) - typically quantified by the transverse thermal conductivity ($K_{zx}$), and involves the measurement of the transverse thermal gradient ($\nabla T_{zx}$) arising due to the Berry curvature of magnon bands under the application of a longitudinal temperature gradient ($\nabla T_{zz}$) and an orthogonal magnetic field ($H$) \cite{Onose297,Watanabe8653,Hirschberger106,PhysRevB.85.134411,PhysRevLett.120.217205,PhysRevLett.115.106603,PhysRevLett.104.066403,PhysRevLett.106.197202}.

Prior experiments of MHE have relied on the use of external chip-thermometers to estimate this gradient across the magnetic specimen \cite{Onose297,PhysRevB.85.134411,PhysRevLett.115.106603,Hirschberger106,Watanabe8653,PhysRevLett.120.217205}. In contrast, we use a device geometry commonly used for the measurement of the Spin Seebeck \cite{PhysRevB.81.214418,Uchida2010,PhysRevLett.124.017203} or Spin Nernst \cite{meyer2017observation,kim2017observation} effect for this purpose. This involves depositing a thin non-magnetic metal (NM) on the specimen surface, and measuring the thermo-electric voltage which develops across this metallic layer as a consequence of the transverse thermal gradient. The specimens are single crystals of Y${_2}$V${_2}$O${_7}$ - a ferromagnetic pyrochlore characterized by a stacking of alternating kagome and triangular sub-lattices along the crystallographic [111] direction. The absence of an inversion center in this structure results in a non-zero Dzyaloshinkii-Moriya interaction which mimics the role of a vector potential - akin to the Lorentz force in the electronic Hall effect \cite{Onose297}. This in turn deflects the thermally driven magnon wave packets, giving rise to the thermal Hall effect of magnons. Our devices are made by depositing a thin ($\approx$10 nm) layer of Pt (or W) on oriented crystal surfaces of Y${_2}$V${_2}$O${_7}$ using dc magnetron sputtering. Measurements are performed with a thermal gradient ($\nabla T_{zz}$) applied across the crystal, and the magnetic field applied along the crystallographic [100] or [111] directions.

\begin{figure*}
	\centering
	\vspace{-0.0cm}
	\hspace{-0.2cm}
	\includegraphics[scale=0.6]{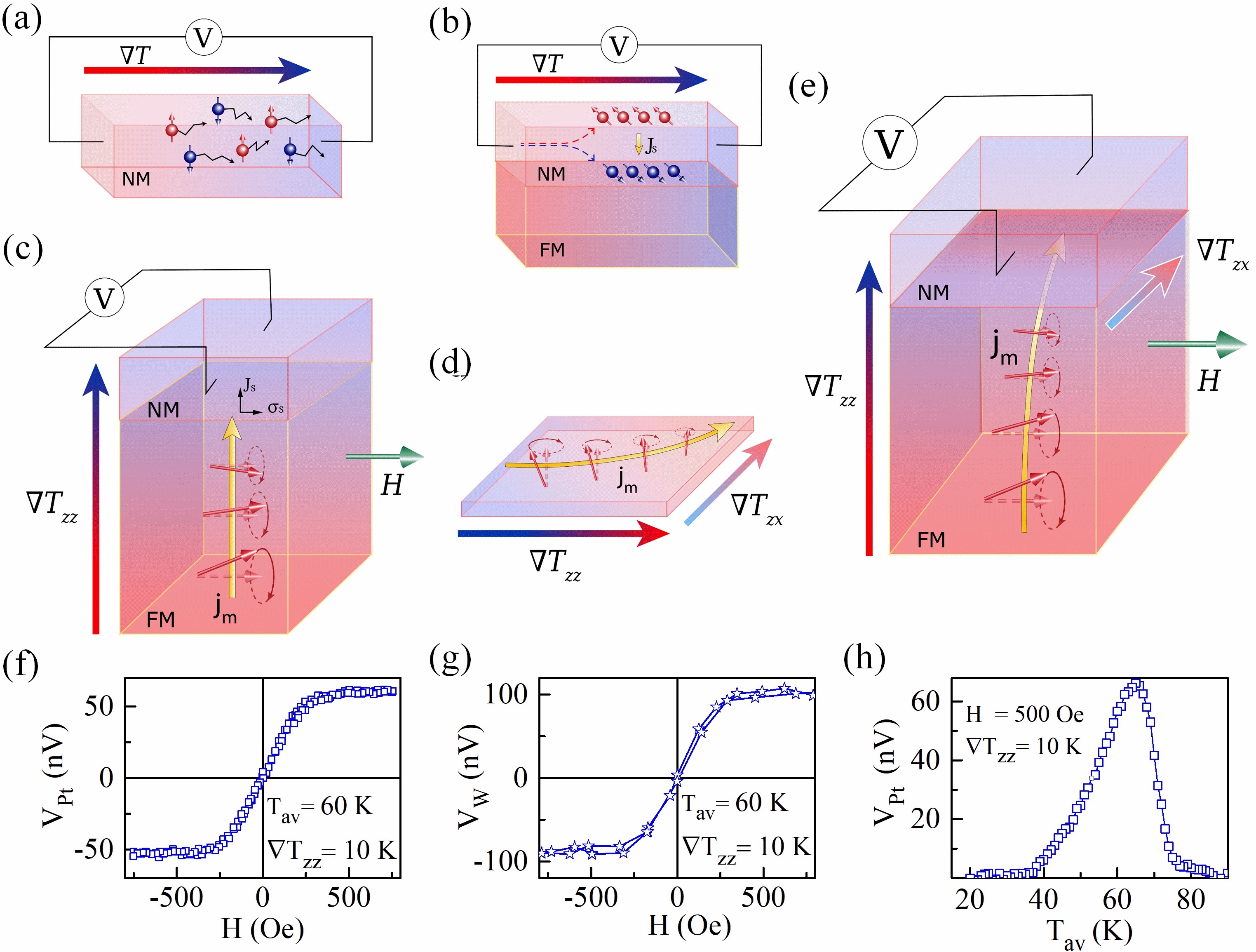}
	\caption{Compendium of possible magneto-thermal effects within our device architecture: (a) A non-magnetic metal (NM) subjected to a temperature gradient ($\nabla T$) generates a thermoelectric voltage as a consequence of the Seebeck effect. If this non-magnetic metal (NM) has large spin orbit coupling, application of a thermal gradient ($\nabla T$) results in a transverse spin current ($j_s$) arising from the spin Nernst (-Ettingshuasen) effect, which is described in (b). When an insulating ferromagnet (FM) is placed adjacent to the NM, a correction to the thermoelectric voltage results as a function of the relative orientation of the spin polarisation of the NM conduction electrons (at the interface) and the magnetization (of the FM layer). (c) depicts the standard geometry utilized for measurements of the longitudinal spin Seebeck effect. Here the application of a temperature gradient across the FM-NM interface could result in a spin current ($j_s$) pumped into the NM, which can then manifest itself as a voltage via the inverse spin Hall effect. This does not appear to contribute significantly to our measured voltage, as described in the supplementary information (SI). (d) depicts the magnon Hall effect, where magnon wave packets traversing under the influence of an applied temperature gradient ($\nabla T_{zz}$) are deflected owing to incipient Dzyaloshinkii-Moriya interactions, resulting in a transverse temperature gradient ($\nabla T_{zx}$). Our device architecture (e) comprises of a  thin NM layer deposited on single crystalline specimens of Y${_2}$V${_2}$O${_7}$. Application of a longitudinal temperature gradient ($\nabla T_{zz}$) results in a transverse temperature gradient ($\nabla T_{zx}$) due to the thermal Hall effect of magnons in the pyrochlore ferromagnet. This transverse gradient is measured via the thermoelectric response of the attached NM layer. The field dependence of this voltage with Pt and W layers is depicted in (f) and (g) respectively, with the magnetic field being applied along the [100] direction of Y${_2}$V${_2}$O${_7}$. (h) depicts the temperature dependence of this voltage as measured in the same configuration.}
\end{figure*} 

The thermo-electric voltage measured across the metallic film could arise from a number of microscopic mechanisms. In addition to the Seebeck effect due to the transverse thermal gradient experienced by the NM, the $\nabla T_{zx}$ could also generate an additional voltage due to the spin Nernst effect (SNE) if the attached NM-layer possesses large spin-orbit interaction \cite{meyer2017observation,kim2017observation}. In addition, a finite contribution from the longitudinal Spin Seebeck Effect cannot be ruled out, if there is spin-pumping across the ferromagnetic-metal interface which can then result in a voltage through the inverse Spin Hall effect \cite{PhysRevB.81.214418,Uchida2010,PhysRevLett.124.017203}. A schematic of our device architecture and the possible contributory mechanisms are summarized in Fig. 1(a)-1(e). The voltage measured across the NM over-layer can be written as:   
\begin{equation} 
V = \left[S + \Delta S_1 + \Delta S_2 (1-m^2_y)\right] \nabla T_{zx} + V_{LSSE}(T)
\end{equation}
where the leading term $S$ is the Seebeck coefficient of the NM-layer, and $\Delta S_1$, $\Delta S_2$ are possible corrections to the Seebeck coefficient due to SNE. With the thickness of Pt layer ($t_{Pt}$) being close to that of the spin diffusion length, the contribution of SNE is expected to be two orders smaller than the bare thermoelectric contribution ($S$). Furthermore, the contribution of $\Delta S_2$ vanishes at fields higher than the magnetic saturation fields (see SI). Fig. 1(f)-1(h) summarizes the field and temperature dependence of the measured voltages in the $H \parallel [100]$ configuration. Below the magnetic transition temperature, the signal is well resolved, and the magnetic field dependence of the voltage follows the magnetization, as is expected for the magnon Hall effect. The fact that the measured voltages have the same sign for both Pt and W layers allows us to rule out a contribution from LSSE $-$ see SI for details. Hence at applied fields larger than the saturation field of Y${_2}$V${_2}$O${_7}$, the measured voltage clearly comprises of the conventional thermoelectric effect in the NM layer alone, and is expected to be a direct measure of the transverse temperature gradient arising due to the magnon Hall effect.

\begin{figure*}
	\centering
	\vspace{-0.0cm}
	\hspace{-0.3cm}
	\includegraphics[scale=0.62]{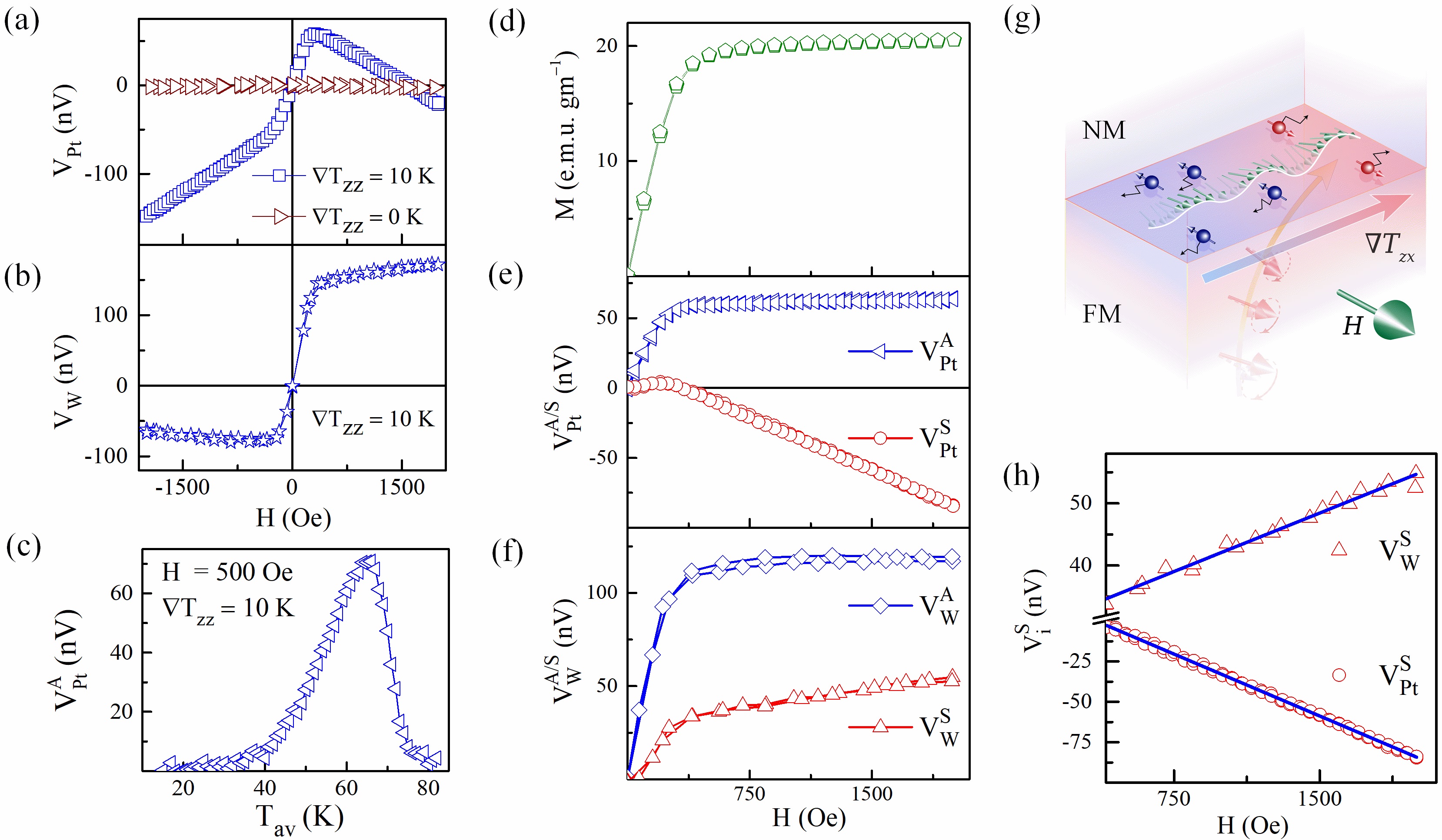}
	\caption{Measurements in the $ H \parallel$[111] configuration and the magnon drag effect: (a) depicts the magnetic field dependence of the measured	voltage as measured across a Pt layer at $T_{av}$ = 60\,K with an applied temperature gradient ($\nabla T_{zz}$) of 10\,K. In contrast to that measured in the $H\parallel[100]$ configuration, the voltage is not anti-symmetric with respect to the applied field, and appears to comprise of an additional symmetric component. In the absence of an applied temperature gradient ($\nabla T_{zz} = 0$), no voltage is detected. (b) depicts the same measurement across a W over-layer, where the (albeit smaller) symmetric contribution is seen to reverse its sign. The temperature dependence of the measured voltage (for Pt) is shown in (c), which matches well with that observed in the $H\parallel[100]$ configuration. The magnetization as measured in the $H\parallel[111]$ direction is shown in (d). The magnetic field dependence of anti-symmetric and symmetric components of the measured voltage as measured across the Pt and W layers are shown in (e) and (f), respectively. The	anti-symmetric component arises from the magnon Hall effect and is in consonance with the sample magnetization (d). The additional symmetric component appears to arise from an interfacial magnon drag effect (g) where a surface spin current in the FM induces a charge current in the adjacent NM layer. The corresponding voltage as measured in the Pt and W layers above the saturation magnetization of Y$_2$V$_2$O$_7$ is shown in (h). The functional form of this voltage is derived from the linear magnetic field dependence of the quenching function L(y), and the sign of its slope is determined by the spin Nernst angle of the NM layer (see main text and SI).}
\end{figure*} 

The temperature (T) dependence this voltage would be determined by the $T$-dependence of both the transverse thermal conductivity ($K_{zx}$) and the Seebeck coefficient of Pt. Since the latter is reported to have a smooth and non-vanishing functional form in the temperature regime of our experiments \cite{doi:10.1063/1.5101028}, the $T$-dependence of the measured voltage is expected to mimic that of the magnon Hall effect. As seen in Fig. 1(h), this voltage rises below the onset of the magnetic transition, exhibits a peak at about 60\,K, and then gradually decreases - in agreement with the thermal Hall conductivity expected to arise due to magnons. A comparison of (i) the magnitude of the transverse temperature gradient arising from the magnon Hall effect, and (ii) the ratio of the transverse and longitudinal thermal conductivities ($\frac{K_{zx}}{K_{zz}}$) as determined from our measurements and that calculated from prior data on Lu${_2}$V${_2}$O${_7}$ \cite{Onose297} reconfirms that the thermoelectric voltage developed across the NM layer is a reliable measure of the magnon Hall effect (see SI for details). A significantly better signal to noise ratio in comparison to earlier reports \cite{Onose297,PhysRevB.85.134411} is observed, attesting to the utility of this device architecture in measurements of the magnon Hall effect. An interesting observation here is that in contrast to what was inferred earlier in Lu$_2$V$_2$O$_7$ \cite{Onose297}, the measured voltage does not appear to vanish as $T \rightarrow 0$, but at a characteristic temperature $T^* \approx 35$\,K.

The field and temperature dependence of the measured voltages in the $H \parallel [111]$ configuration are summarized in Fig.2(a)-2(c). The contrast with the field-dependent data obtained with $H \parallel [100]$ is significant - since the antisymmetric voltage arising due to the magnon Hall effect is now seen to be superimposed with an additional  \emph{symmetric} component. This is evident in Fig.2(e)-2(f), where the field dependence of the symmetric and antisymmetric components are individually depicted. The antisymmetric component of the measured voltage saturates in consonance with the magnetization (shown in Fig. 2(d)), which is agreement with magnon Hall effect. In contrast, the symmetric component exhibits a significant field dependence. In the low magnetic field regime, this could potentially arise due to magneto-resistive (MR) effects such as anisotropic magneto-thermal resistance \cite{PhysRevB.87.094409}, spin-orbit magnetoresistance \cite{Zhoueaao3318}, and the spin Hall magnetoresistance \cite{PhysRevLett.110.206601}. However, all these MR effects necessarily saturate with the magnetization of Y${_2}$V${_2}$O${_7}$, implying that they cannot drive this symmetric component above the saturation magnetization fields.

A linearly varying symmetric component beyond the saturation magnetization can arise due to the magnon drag effect (MDE) \cite{PhysRevB.13.2072,doi:10.1063/1.3672207,PhysRevB.94.144407,costache2012magnon,PhysRevB.99.094425}, in which an in-plane magnon current at the interface could drag the conduction electrons in the adjacent NM layer (depicted in Fig.2 (g)). This advective contribution to the (otherwise diffusive) thermopower is well known in magnetic metals, and has been described using the hydrodynamic \cite{PhysRevB.13.2072,costache2012magnon} and relativistic spin motive force \cite{PhysRevB.94.144407,doi:10.1063/1.3672207,PhysRevB.99.094425} approaches. The former models a magnetic metal as a two-component mixture of interacting magnons and electrons, whereas the latter relies on a spin-orbit interaction driven conversion of a magnon heat flux to an electric current \cite{PhysRevB.99.094425,doi:10.1063/1.3672207,PhysRevB.94.144407}. The electric current ($\vv{j_d}$) due to the MDE can be expressed as \cite{PhysRevB.99.094425,doi:10.1063/1.3672207,PhysRevB.94.144407}: 
\begin{equation}
\vv{j_d} = \frac{\sigma_{\uparrow}-\sigma_{\downarrow}}{e^2} \braket{\vv{F_i}} = \beta \sigma P_s \frac{\hbar}{2e} \frac{\vv{j^q_m}}{sD}
\end{equation}
Here, $\sigma$ is the electronic conductivity at the interface ($\sigma =\sigma_{\uparrow}+ \sigma_{\uparrow}$, with $\sigma_{\downarrow}$ and $\sigma_{\uparrow}$ being the conductivities with up and down spin-polarizations respectively), and $P_s =\frac{\sigma_{\uparrow}-\sigma_{\downarrow}}{\sigma}$ is the effective spin-polarization. $\beta$ is a dissipative spin-transfer torque parameter, $e$ is the charge of an electron, $s$ is the magnon density at the interface, and $D$ is the spin stiffness. $\vv{j^q_m} = k_m \nabla T_{zx}$ is the magnon heat current, with  $k_m$ being the thermal conductivity of magnons at the interface. This electric current ($\vv{j_d}$) would be symmetric with respect to the applied field, since both $P_s$ and $\nabla T_{zx}$ reverse their polarities with $\vv{H}$.

The magnetic-field dependence of this drag is determined by that of the magnon heat flux, $j^q_m(H)$, which can be expressed in terms of the quenching function ($L(y)$). This function varies approximately linearly with the strength of the effective magnetic field $\lvert B _{eff}\rvert $, in the low magnetic field limit. Since $B_{eff}$ is determined by both the external magnetic field ($H$), and the sample magnetization ($M$), $L(y)$ is thus expected to vary linearly with $\vert H \vert$ above the saturation field of Y${_2}$V${_2}$O${_7}$. Fig.2(h) shows the magnetic field dependence of the symmetric magnon drag contribution as measured in the $H \parallel [111]$ configuration, where this linearity is clearly seen. Moreover, the polarity of this drag is also observed to reverse when the NM layer is changed from Pt to W. Since the sign of $P_s$ at the interface is determined by the SNE of the attached NM layer, the spin Nernst angle ($\theta_{SN}$) of that NM layer could alter the polarity of the dragged voltage. That the slope of this dragged voltage in the saturated magnetization regime changes its sign when the W layer replaces the Pt layer in the $H \parallel [111]$ configuration (Fig. 2(h)) is a consequence of the fact that the sign of $\theta_{SN}$ in Pt is opposite to that in W \cite{kim2017observation}, (see SI for details).

The microscopic reason for the existence of a magnon drag contribution to the measured thermopower in the $H \parallel [111]$ configuration appears to be the non-trivial magnon topology in this class of materials. Theoretical calculations suggest that the magnon-band structure of these pyrochlore ferromagnets comprises of four distinct bulk-bands, with all of them possessing a  non-vanishing Berry curvature \cite{PhysRevLett.117.157204}. This magnon band structure is also sensitive to the direction of the applied field, where a band-gap between the first (lowest-energy) band and the second band exists as long as the field is \emph{not within} the $\{100\}$ plane of the crystal \cite{PhysRevLett.117.157204}. More importantly, a topologically protected surface magnon state exists between the first and second bulk bands as long as this gap opening condition is satisfied. Since the magnon drag is linked to the momentum-transfer and spin-transfer processes (the $\beta$ factor in Equation 2), we infer that an enhanced electron-magnon interaction at the interface \cite{PhysRevB.99.094425,doi:10.1063/1.3672207,PhysRevB.94.144407} is directly responsible for this additional voltage observed in the $H \parallel [111]$ configuration.

The relative dearth of experimental probes has hampered the exploration of topologically protected magnon surface states in strongly correlated magnets. Inelastic neutron scattering - the measurement of choice in mapping magnon band topology - is inherently a bulk technique, with limited sensitivity to surface states \cite{PhysRevLett.113.047202}. Brillioun light scattering could be more viable, having been used earlier to identify spin waves propagating along the surface of magnetic crystal \cite{PhysRevLett.39.1561}. However it remains to be demonstrated whether the momentum transfer window accessible in this technique would enable an unambiguous determination of these magnon surface states. Our experiments show that the thermal and electronic manifestations of magnon transport can provide a viable alternative. More complex device architectures can be envisaged to identify and even possibly manipulate these surface magnon states. For instance,  it has been suggested that the ferromagnetic pyrochlores could also harbor Weyl magnons owing to a linear crossing between the second and the third bulk bands. This could manifest itself as a magnon arc - in analogy to the Fermi arcs in electronic topological systems - and could have implications in dissipation-less spintronic devices.

A.D. acknowledges UGC (Govt. of India) for a Senior Research Fellowship. A.D. and S.N. acknowledge funding support by the Department of Science and Technology (Govt. of India) under the DST Nanomission Thematic Unit Program (SR/NM/TP13/2016), and a Science and Engineering Research Board grant (SPR/2020/389). D.P acknowledges support from the Oxford-ShanghaiTech collaboration project and the UK Engineering and Physical Sciences Research Council (grant no. EP/T028637/1)..

\bibliography{Bibliography-1}

\end{document}